\pdfoutput=1 
\documentclass[type=preprint,linenumbering=off]{sphenix}
\usepackage{amsmath}
\usepackage{amssymb}     
\usepackage{cleveref}
\usepackage{mathrsfs}
\usepackage{xspace}
\usepackage{xcolor}      
\usepackage{booktabs}    
\usepackage[percent]{overpic}
\crefrangeformat{figure}{Figs. #3#1#4--#5#2#6}

\usepackage{orcidlink}

\usepackage{defs}

\providecommand{\wetacogx}{\ensuremath{w_{\eta}}\xspace}

\definecolor{p9blue}{rgb}{0.0,0.3,0.85}
\definecolor{p9red}{rgb}{0.65,0.1,0.1}

\def\GeV{\ifmmode {\mathrm{\ Ge\kern -0.1em V}}\else
                   \textrm{Ge\kern -0.1em V}\fi \xspace}%

\def\MeV{\ifmmode {\mathrm{\ Me\kern -0.1em V}}\else
                   \textrm{Me\kern -0.1em V}\fi \xspace}%

\def\TeV{\ifmmode {\mathrm{\ Te\kern -0.1em V}}\else
                   \textrm{Te\kern -0.1em V}\fi \xspace}%

\begin{document}

\renewcommand{\thefootnote}{$\star$}

\title{Measurement of isolated prompt photon production in $p{+}p$ collisions at $\sqrt{s}=200~\mathrm{GeV}$ with the sPHENIX detector}
\author{sPHENIX Collaboration\footnote{See the appendix for the list of collaboration members}}
\date{\today}


\doctag{sPH-JET-2026-02}
\docdoi{-}


\maketitle
\begin{abstract}
The differential cross section of isolated prompt photon production is measured as a function of photon transverse energy ($E_{\mathrm{T}}^{\gamma}$) in proton--proton ($p{+}p$) collisions at $\sqrt{s}=200~\mathrm{GeV}$. The data were recorded in $2024$ with the sPHENIX detector at the Relativistic Heavy Ion Collider.
Photons are reconstructed in $|\eta^{\gamma}|<0.7$ and $12<E_{\mathrm{T}}^{\gamma}<32~\mathrm{GeV}$ using the electromagnetic calorimeter, and an isolation requirement is imposed using both the electromagnetic and hadronic calorimeters.
The measured cross section is compared with the \textsc{Pythia} Monte Carlo event generator and perturbative quantum chromodynamics (pQCD) calculations at next-to-leading and next-to-next-to-leading order. The pQCD calculations are consistent with the result within the quoted uncertainties. This measurement provides a test of pQCD calculations for a process with sensitivity to the gluon parton distribution function of the proton and establishes the $p{+}p$ baseline for forthcoming sPHENIX measurements of isolated prompt photons in heavy-ion collisions.

\end{abstract}

\section{Introduction}
\label{sec:introduction}

Prompt photons are defined as those produced either directly in the short-distance partonic hard scattering, referred to as \textit{direct photons}, or through the collinear fragmentation of a showering parton, referred to as \textit{fragmentation photons}. The production of prompt photons is calculable within the framework of perturbative quantum chromodynamics (pQCD). At leading order (LO), direct photons are produced predominantly through quark--gluon Compton scattering ($qg \to q\gamma$) and quark--antiquark annihilation ($q\bar{q} \to g\gamma$). At next-to-leading order (NLO), the calculation includes higher-order corrections to the direct component as well as contributions in which the observed photon arises from the collinear fragmentation of a final-state parton~\cite{Aurenche:2006vj}. The separation between the direct and fragmentation components is factorization-scheme dependent, and only the sum of the two components is physical~\cite{Gehrmann:2022cih}.

To reduce the contribution of fragmentation photons and background photons from hadron decays, an isolation requirement can be imposed. Such a requirement keeps the cross section infrared-safe and perturbatively calculable. This is performed by restricting the amount of transverse energy surrounding the photon within a fixed cone in pseudorapidity ($\eta$) -- azimuthal angle ($\phi$) space. Prompt photon cross sections have been measured at fixed-target experiments over a wide range of collision energies~\cite{WA70:1987vvj,UA6:1998ozg,FermilabE706:2004emk}. Isolated prompt photon cross sections have been measured at the Tevatron~\cite{D0:2005ofv,CDF:2009rtv} in $p$+$\bar{p}$ collisions and extensively at the Large Hadron Collider (LHC) in \pp collisions~\cite{CMS:2018qao, ATLAS:2016fta, ATLAS:2013sdn, ATLAS:2011ezy, ALICE:2019rtd, ALICE:2024kgy} as well as in \ppb~\cite{ALICE:2025bnc,ATLAS:2019ery} and \pbpb~\cite{CMS:2020oen,ALICE:2024yvg,ATLAS:2015rlt} collisions across various collision energies. At the Relativistic Heavy Ion Collider (RHIC)~\cite{HARRISON2003235}, prompt photon cross sections without an isolation requirement have been measured by PHENIX in \pp~\cite{PHENIX:2012jgx}, $d{+}\mathrm{Au}$~\cite{PHENIX:2012krx}, and $\mathrm{Au}{+}\mathrm{Au}$~\cite{PHENIX:2005yls} collisions, and by STAR in \pp{} and $d{+}\mathrm{Au}$~\cite{STAR:2009ojw}. An isolated direct-photon cross section at higher RHIC energy has been measured by PHENIX in \pp{} at $\sqrt{s} = 510$~GeV~\cite{PHENIX:2022lgn}. The LHC measurements use a data-driven double-sideband technique to extract the prompt-photon signal, while the previous RHIC measurements at $\sqrt{s} = 200$~GeV relied on statistical subtraction of the calculated decay-photon yield.

In \pp collisions, the prompt photon production provides a stringent test of pQCD predictions. In addition, the quark--gluon Compton process gives the dominant contribution to prompt photon production~\cite{PHENIX:2010vgy}, so that the cross section is particularly sensitive to the gluon parton distribution function (PDF) of the proton~\cite{Ichou:2010wc}. These measurements in \pp{} collisions also provide a crucial reference for identifying possible modifications in the nuclear medium, since prompt photons in heavy-ion collisions do not strongly interact with the final-state QGP and their nuclear modification factor is directly sensitive to nPDFs~\cite{Klasen:2023uqj} and initial-state effects~\cite{Vitev:2008vk}.

This paper presents the sPHENIX measurement of the differential cross section of isolated prompt photon production as a function of \etg{} in \pp{} collisions at \comHEP{}, in the kinematic range $|\etag|<0.7$ and $12<\etg<32\GeV$. The measurement uses $\mathscr{L}=64.4\,\pb$ from the data sample collected in \RunOne. After applying the reconstructed vertex requirement described in Sec.~\ref{sec:data_sim}, the effective luminosity of the selected event sample is approximately $49\,\pb$. The measurement is reported with a particle-level isolation requirement that the total transverse energy (\isoET) of final-state particles within $\DR=0.3$ around the photon, excluding the photon itself, is below $4~\GeV{}$. The cross section is compared to NLO pQCD calculations from \jetphox~\cite{Aurenche:2006vj}, the NLO pQCD calculation by Vogelsang~\cite{Gordon:1993qc}, the next-to-next-to-leading-order (NNLO) pQCD calculations from \nnlojet{}~\cite{NNLOJET:2025rno}, the \pythia{}~$8.307$~\cite{Sjostrand:2014zea} Monte Carlo (MC) generator with the Detroit tune~\cite{Aguilar:2021sfa}, and the previous inclusive prompt photon measurement by PHENIX~\cite{PHENIX:2012jgx}.

\section{sPHENIX detector}
\label{sec:sphenix_info}

sPHENIX~\cite{PHENIX:2015siv, Belmont:2023fau} is a new detector designed to measure jet and heavy-flavor probes of the quark-gluon plasma created in Au+Au collisions at RHIC. A precision tracking system enables measurements of heavy-flavor and jet-substructure observables, while the electromagnetic and hadronic calorimeter system is crucial for measuring the energy of jets and identifying photons and electrons. For the present measurement, only the electromagnetic and hadronic calorimeters, as well as a forward minimum bias detector (MBD) are used. A full description of the apparatus is given in Ref.~\cite{Sphenix:TDR}.

The Electromagnetic Calorimeter (EMCal) is a scintillating-fiber tungsten sampling calorimeter~\cite{Aidala:2020toz,sPHENIX:2017lqb,Yu:2024quo} with an approximately two-dimensional projective geometry in $\eta$-$\phi$ and approximately $0.025 \times 0.025$ segmentation. The EMCal corresponds to approximately $20$ radiation lengths and $0.8$ nuclear interaction lengths at $\eta = 0$, and its $\pi^0/\eta\rightarrow\gamma\gamma$ \textit{in-situ} energy calibration is described in Ref.~\cite{sPHENIX:2025dET}. The scintillation light from each tower is collected via a light guide and detected by four silicon photomultipliers. 

The Inner Hadronic Calorimeter (IHCal) and Outer Hadronic Calorimeter (OHCal)~\cite{Belmont:2023fau,sPHENIX:2017lqb}, together with the EMCal, provide approximately five nuclear interaction lengths at $\eta=0$, with aluminum (IHCal) and steel (OHCal) absorber plates interleaved with scintillating tiles and $\Delta\eta\times\Delta\phi = 0.1\times 0.1$ segmentation. The IHCal and OHCal are used in this measurement to compute the photon isolation transverse energy together with the EMCal. The calorimeter data acquisition system for the EMCal, IHCal and OHCal digitizes the photomultiplier signals six times per bunch crossing. Two of these digitized samples are utilized in a digital triggering scheme which selects events up to $15$~kHz for recording.
The MBD, located at $3.51 < |\eta| < 4.61$, provides the primary reconstructed collision vertex from the timing difference between the two sides of the detector relative to the collision region.

\section{Data and simulation}
\label{sec:data_sim}

\subsection{Event Selection}
\label{ssec:eventsel}
\RunOne{} recorded \pp{} collisions in two beam-crossing-angle configurations, a $0$-mrad period and a $1.5$-mrad period, which differed substantially in instantaneous luminosity.
At the instantaneous luminosities delivered during \RunOne{}, a non-negligible fraction of bunch crossings contain more than one inelastic \pp{} interaction, with mean per-bunch-crossing interaction multiplicities of $\mu \approx 0.24$ at $0$~mrad and $\mu \approx 0.08$ at $1.5$~mrad. This effect, referred to as pile-up, can bias the primary vertex reconstructed by the MBD, which in turn affects the reconstructed cluster kinematics and shower shape variable (defined in Sec.~\ref{sec:analysis}) distributions of photon candidates.

Events are selected by a single high-\et photon trigger requiring the energy sum in any $8\times 8$ EMCal tower window to exceed $4$~\GeV{}. The trigger efficiency reaches at least $99.5\%$ for $\etg > 8.5$~\GeV{}, and the residual per-bin turn-on is absorbed into the photon-trigger efficiency correction. Events are further required to have a reconstructed MBD $z$-vertex in $|\vz|<60~\text{cm}$, retaining approximately $76\%$ of the recorded events.

\subsection{Monte Carlo Simulations}
\label{ssec:mc}
Simulations of prompt photon and inclusive jet MC events are generated with \pythia~$8.307$~\cite{Sjostrand:2014zea} using the Detroit tune~\cite{Aguilar:2021sfa}. In every sample, both \texttt{HardQCD:all} and \texttt{PromptPhoton:all} are enabled. The \texttt{PromptPhoton:all} processes generate direct photons produced in the hard scattering, while the \texttt{HardQCD:all} processes provide fragmentation photons from final-state partons. To ensure sufficient statistics at high transverse momentum, multiple samples are generated with different partonic hard-scattering thresholds, $\hat{p}_{\rm T,\min}$. The prompt photon samples are filtered by requiring a truth photon, while the inclusive jet samples are filtered by requiring a truth jet, in both cases within $|\eta|<1.5$. Thus, the jet samples contain all the major contributions to high-\et clusters, making them inclusive. The samples are combined using their effective cross sections to produce a continuous truth-\etg{} spectrum across the sample boundaries. 
Prompt photon samples are also generated with the \herwig{}~$7.3$ generator~\cite{Bellm:2015jjp} using the Nashville tune~\cite{Qureshi:2024eqz} as a generator cross-check on the isolation efficiency. 

The generated \pythia{} and \herwig events are propagated through the full sPHENIX detector using the \geant simulation package~\cite{AGOSTINELLI2003250} with simulated noise in the calorimeter towers configured to match the noise in data. The simulation includes EMCal response effects associated with electromagnetic shower leakage, SiPM pixel occupancy, and fiber attenuation. Photons are then reconstructed following the same procedure as in the data. To account for differences between data and simulation in the EMCal energy resolution, an additional \etg{}-dependent Gaussian smearing, $\sigma_{+MC}$, of approximately $2$\% at high \etg{} is applied to the simulated cluster \et{}, with a width chosen such that the energy resolution matches the one measured in data via $\pi^0$ and $\eta$ mass-peak fits. 

At the truth level, signal photons are defined as prompt photons that satisfy the isolation requirement: the total transverse energy of all final-state particles within $\DR=0.3$, excluding the photon itself, is below $4$~\GeV. A reconstructed photon candidate is matched to a generated photon when that photon contributes the largest fraction of the simulated energy deposited in the candidate's EMCal cluster.

To account for pile-up (PU) effects, dedicated full \geant{} pile-up MC simulation samples are generated, in which each event combines a hard \pythia{} event with a minimum-bias \pythia{} event placed at an independent generated vertex along the beam line. Among events with at least one high-energy EMCal cluster, the fractions with two or more inelastic \pp{} interactions are $f_{\mathrm{PU}}^{0\,\mathrm{mrad}}=0.22$ and $f_{\mathrm{PU}}^{1.5\,\mathrm{mrad}}=0.079$, computed from the per-bunch-crossing interaction multiplicities, $\mu$, corresponding to $0$ and $1.5$ $\mathrm{mrad}$, respectively. The simulation sample used in this analysis is constructed within each crossing-angle period by combining the single-interaction and pile-up samples weighted according to $(1-f_{\mathrm{PU}})$ and $f_{\mathrm{PU}}$ respectively, and the two crossing-angle periods are then combined according to their integrated luminosities. A per-event truth-vertex reweighting is also applied so that the reconstructed-vertex distribution in simulation matches the data after event selection. The pile-up description as well as the vertex reweighting are required for adequate modeling of shower shape (described in Sec.~\ref{sec:analysis}) distributions and other aspects of the EMCal performance in data. The analysis was also carried out separately for the $0$ and $1.5$ $\mathrm{mrad}$ data sets and the results were consistent.

Prompt photon MC samples are used for efficiency corrections, unfolding, and the purity signal-leakage correction. Inclusive jet MC samples are used to optimize the photon identification (\gid).

\section{Analysis}
\label{sec:analysis}

\subsection{Photon Reconstruction and Identification}
\label{ssec:reco}
Photon candidates are reconstructed by clustering EMCal towers. Superclusters are formed by grouping contiguous towers with energies above $70$~\MeV{}, with a total energy exceeding $0.5$~\GeV{}. Superclusters are then divided into sub-clusters using a local peak-finding algorithm within a $3\times 3$ tower grid~\cite{David:2000wfa}. Reconstructed photons have an energy resolution $\sigma_E/E$ of approximately $7$\% at $\etg \approx 12~\GeV$ and approximately $6$\% at $\etg \approx 32~\GeV$.

The shapes of the measured energy deposits in the EMCal, referred to as shower shape, are used at several stages of the analysis. Simple selections on shower shape variables are first applied to reject residual electronic noise and to reduce correlations between photon identification and isolation. A boosted decision tree (BDT) classifier is then used to reject non-collision backgrounds (NCB), which have shower profiles and timing distributions distinct from those of clusters produced in collisions. Finally, a separate photon identification BDT is used to discriminate prompt photon candidates from background clusters, primarily from merged photons produced in neutral meson decays. The shower shape variables and the corresponding selections are described below.

\subsubsection*{Shower shape variables}
A set of shower shape variables is constructed from the EMCal tower energies in each reconstructed cluster. These variables characterize the transverse size, asymmetry, and energy concentration of the measured energy deposit. They are used for three purposes in this analysis: the pre-selection, the NCB BDT, and the photon identification BDT. The photon identification BDT uses eight variables: two transverse second-moment widths $w_\eta$ and $w_\phi$ in the cluster center-of-gravity frame, four asymmetry observables $E_{\mathrm{t}1}$--$E_{\mathrm{t}4}$ built from the $2\times 2$ tower block around the center of gravity, and two energy-fraction ratios $E_{1\times 1}/E_{3\times 3}$ and $E_{3\times 2}/E_{3\times 5}$. The NCB classifier described below uses these eight variables together with an extended set of shower shape observables: radial second-moment widths $w_{n\times 2}$ in $n\times 2$ tower blocks for $n=3,5,7$, central-tower energy fractions $E_{1\times 1}/E_{1\times n}$, $E_{1\times 1}/E_{m\times 1}$, and $E_{1\times 1}/E_{2\times 2}$, and $2\times 2$-block energy fractions $E_{2\times 2}/E_{m\times n}$. All variables are defined in Table~\ref{tab:bdt_features}. 
Two variables of particularly high discriminating power, the transverse second-moment widths $w_\eta$ and $w_\phi$ are given explicitly by
\begin{equation}
w_\alpha = \sqrt{\frac{\sum_i E_i\,(\alpha_i-\bar\alpha)^2}{\sum_i E_i}},
\qquad
\bar\alpha = \frac{\sum_i E_i\,\alpha_i}{\sum_i E_i},
\qquad \alpha=\eta,\phi,
\label{eq:weta_wphi}
\end{equation}
where the sum runs over towers $i$ in the cluster.



\begin{table}[hbtp!]
    \centering
    \begin{tabular}{c|p{0.78\textwidth}}
        Feature & Definition \\
        \hline
        \multicolumn{2}{l}{\textit{Transverse second moment widths}} \\
        \hline
        $w_\eta$ & Energy-weighted second moment of $\eta$ in the cluster center-of-gravity frame \\
        $w_\phi$ & Energy-weighted second moment of $\phi$ in the cluster center-of-gravity frame \\
        \multicolumn{2}{l}{\textit{$2\times 2$ block asymmetry observables}} \\
        $E_{\mathrm{t}1}$ & $(E_1+E_2+E_3+E_4)/E_\mathrm{tot}$, fraction of cluster energy in the $2\times 2$ block around the cluster center of gravity (CoG); $E_1$ is the CoG tower, $E_2$ the tower adjacent in $\phi$, $E_3$ the diagonal-corner tower, and $E_4$ the tower adjacent in $\eta$ \\
        $E_{\mathrm{t}2}$ & $(E_1+E_2-E_3-E_4)/E_\mathrm{tot}$, $\eta$-direction asymmetry of the $2\times 2$ block \\
        $E_{\mathrm{t}3}$ & $(E_1-E_2-E_3+E_4)/E_\mathrm{tot}$, $\phi$-direction asymmetry of the $2\times 2$ block \\
        $E_{\mathrm{t}4}$ & $E_3/E_\mathrm{tot}$, diagonal-corner tower fraction \\
        \hline
        \multicolumn{2}{l}{\textit{Energy fraction ratios}} \\
        \hline
        $E_{1\times 1}/E_{3\times 3}$ & Ratio of central tower energy to the surrounding $3\times 3$ tower energy \\
        $E_{3\times 2}/E_{3\times 5}$ & Ratio of energy in a $3\times 2$ ($\eta\times\phi$) tower region to a $3\times 5$ region around the cluster center of gravity \\
        \hline
        \hline
        \multicolumn{2}{l}{\textit{Radial second moment widths}} \\
        \hline
        $w_{n\times 2}$ & Radial second-moment widths in $n\times 2$ tower blocks around the cluster center of gravity, $n=3,5,7$ \\
        \hline
        \multicolumn{2}{l}{\textit{Central-tower energy fractions}} \\
        \hline
        $E_{1\times 1}/E_{1\times n}$ & Central-tower energy fraction relative to $1\times n$ ($\eta\times\phi$) tower strips, $n=3,5,7$  \\
        $E_{1\times 1}/E_{m\times 1}$ & Central-tower energy fraction relative to $m\times 1$ ($\eta\times\phi$) tower strips, $m=3,5,7$ \\
        $E_{1\times 1}/E_{2\times 2}$ & Central-tower energy fraction relative to the $2\times 2$ tower block \\
        \hline
        \multicolumn{2}{l}{\textit{Extended energy fraction ratios}} \\
        \hline
        $E_{2\times 2}/E_{m\times n}$ & $2\times 2$-block energy fraction relative to extended $m\times n$ tower regions, with $(m,n)\in\{(3,3),(3,5),(3,7),(5,3)\}$ \\
    \end{tabular}
    \caption{Shower shape variables used in this analysis. The variables above the double horizontal line are used as input features for the photon identification BDT. All variables in the table are used as input features for the NCB BDT.}
    \label{tab:bdt_features}
\end{table}

\subsubsection*{Non-collision background removal} NCB clusters originate predominantly from beam--pipe interactions upstream of the interaction region. The resulting background particles propagate approximately parallel to the beam direction, entering the EMCal at shallow angles relative to the sPHENIX detector $z$-axis. They fire the photon trigger at a rate that is non-negligible relative to the physics signal in the kinematic range of this analysis. NCB clusters are characterized by two main features: broad energy deposits in $\eta$, which produce large tails in the $w_\eta$ distribution, and EMCal cluster--MBD time differences shifted toward negative values. In contrast, clusters from particles produced in the collisions (\textit{physics clusters}) have narrower electromagnetic shower profiles and cluster--MBD time differences centered near zero.
A multivariate NCB classifier is built as a BDT using the \textsc{XGBoost} package~\cite{Chen:2016xgboost}. The BDT is trained to distinguish a NCB-enriched data sample from an inclusive-MC sample  of physics clusters, using the shower shape variables listed in Table~\ref{tab:bdt_features}, including the additional shower shape observables listed below the double horizontal line in the same table. The NCB-enriched sample is selected in data using clusters with negative cluster--MBD time differences and no back-to-back recoil jet~\cite{sPHENIX:2026jgh}. To reduce sensitivity to differences in the kinematic distributions of the two training samples, the samples are reweighted to have a flat cluster-\et prior, while cluster \et, $\eta$, and $z_\mathrm{vertex}$ are included together with the shower shape variables as input features. The classifier output, referred to as the NCB BDT score, is required to exceed $0.5$, retaining approximately $99\%$ of physics clusters with ${\sim}99\%$ purity.

\subsubsection*{Pre-selection}
In addition to the NCB BDT cut, three rectangular cuts are imposed on the shower shape variables: $E_{1\times 1}/E_{3\times 3} < 0.98$, $0.6 < E_{\mathrm{t}1} < 1.0$, and $0.8 < E_{3\times 2}/E_{3\times 5} < 1.0$. 
This pre-selection serves two purposes. First, it rejects residual electronic noise and other poorly reconstructed clusters that populate extreme regions of shower shape phase space. Second, it removes regions where the photon identification BDT score is strongly correlated with the reconstructed isolation energy. Reducing this correlation improves the independence of the identification and isolation axes, which is required for the double-sideband method described in Sec.~\ref{ssec:abcd}.

\begin{figure}[tbp!]
    \centering
    \includegraphics[width=0.45\linewidth]{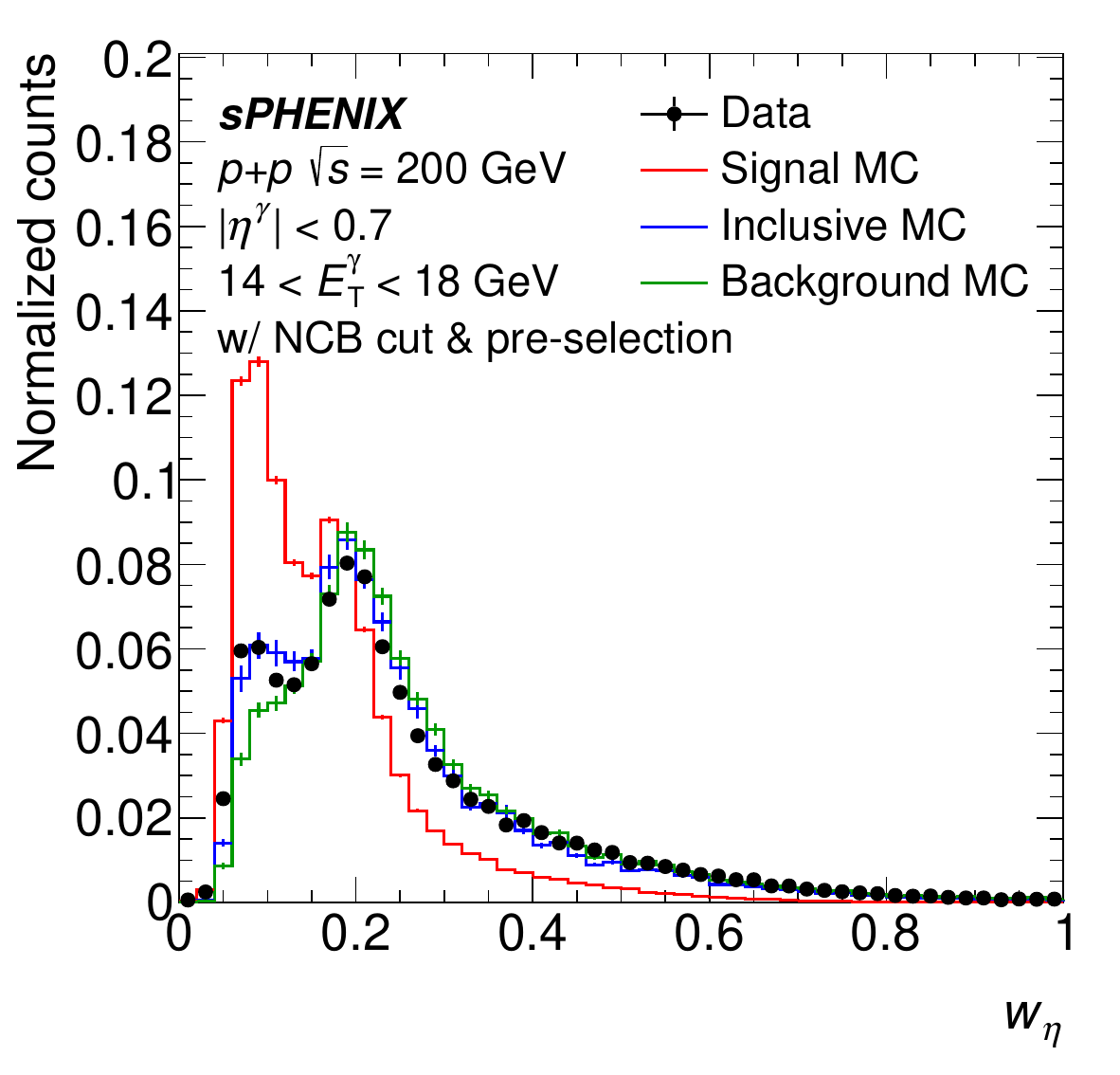}
    \includegraphics[width=0.45\linewidth]{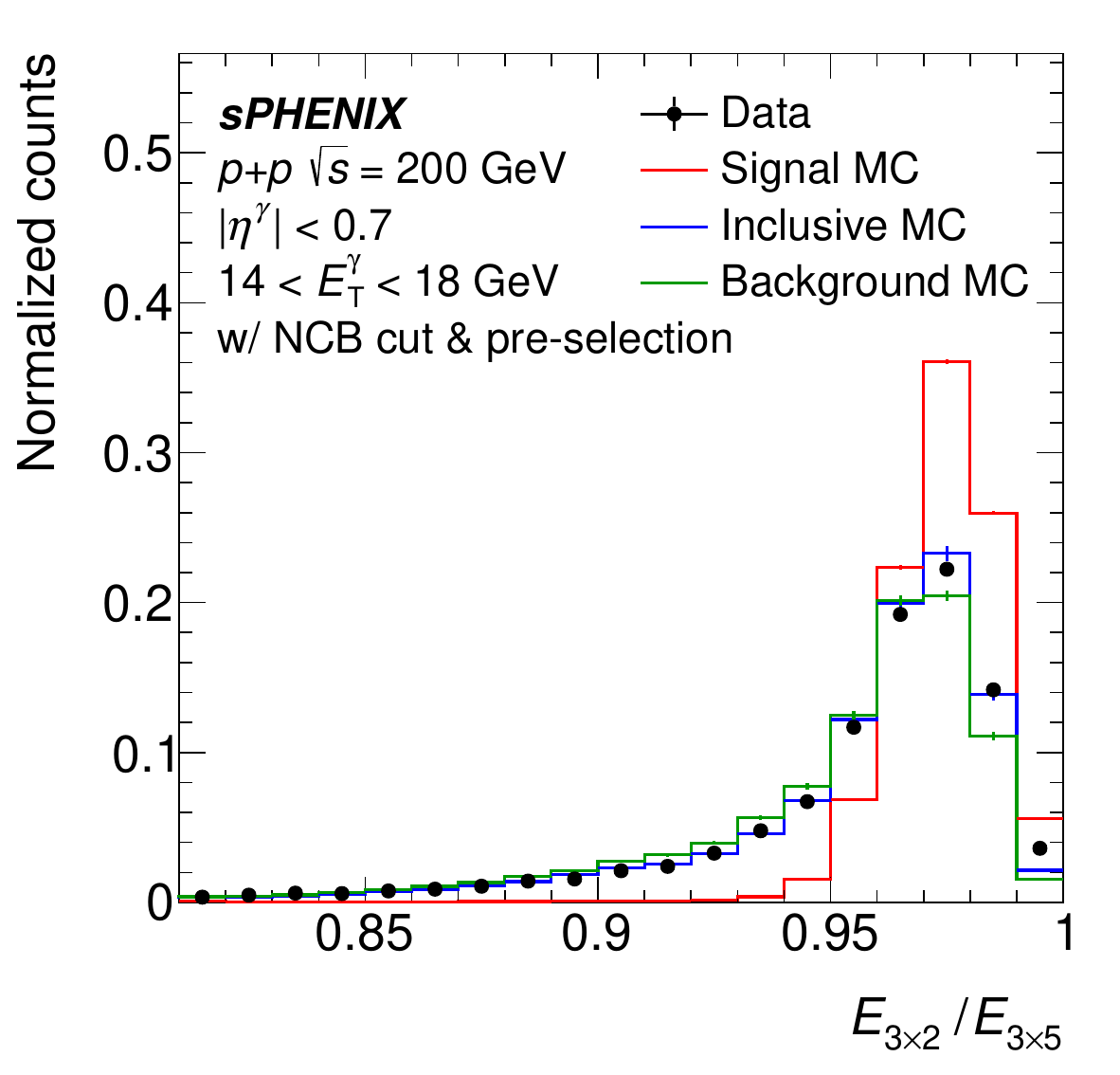}
    \caption{Distributions of the shower shape variables \wetacogx{} (left) and $E_{3\times 2}/E_{3\times 5}$ (right) after the non-collision background (NCB) and pre-selection requirements, before applying the photon identification selection, for $14<\etg<18~\GeV$ and $|\etag|<0.7$. Black points show data, the red histogram shows the prompt photon signal MC, the blue histogram shows the inclusive MC, and the green histogram shows the background MC. The data and MC distributions are normalized such that the integral is unity.}  \label{fig:showershape_dis}
\end{figure}

Two representative shower shape distributions are shown in Figure~\ref{fig:showershape_dis}, where data are compared with the prompt photon signal MC and the inclusive MC after the NCB and pre-selection cuts, for $14<\etg<18~\GeV$. Signal photons in MC preferentially populate the narrow shower regions of \wetacogx{} and $E_{3\times 2}/E_{3\times 5}$, while the data remain dominated by decay-photon backgrounds at the pre-selection level and therefore more closely follow the inclusive MC distributions.

\subsubsection*{Photon identification}

Prompt photons typically produce narrow, single-core electromagnetic showers in the EMCal. The dominant background arises from merged two-photon clusters from $\pi^0\to\gamma\gamma$ and other neutral meson decays, with a smaller contribution from the early stages of other hadron's showers. These backgrounds tend to produce broader or more asymmetric energy profiles. The two photons from $\pi^0\to\gamma\gamma$ start to merge into a single EMCal cluster above $p_T\approx5~\GeV$, and for $\eta\to\gamma\gamma$ above $p_T\approx20~\GeV$; the merging is gradual rather than a hard threshold. Shower shape variables are used to discriminate prompt photons from background photons.

After the NCB cut and pre-selection, photon candidates are classified using a second BDT trained to separate truth-matched photon clusters in the \pythia{} samples from background clusters in the \pythia{} inclusive samples. The eight shower shape variables listed in Table~\ref{tab:bdt_features} are used as training features, together with three kinematic variables: cluster \et, cluster $\eta$, and the collision vertex position $\vz$. Class reweighting is applied to balance the signal and background yields, and inverse-PDF weights are used to flatten the \pt and $\eta$ distributions, so that the BDT learns the shower shape discrimination uniformly across the analysis kinematic range. The dominant input feature is $w_\eta$ (Eq.~\ref{eq:weta_wphi}), the energy-weighted second moment of the tower $\eta$ distribution evaluated in the cluster center-of-gravity reference frame.

\begin{figure}[tbp!]
    \centering
    \includegraphics[width=0.55\linewidth]{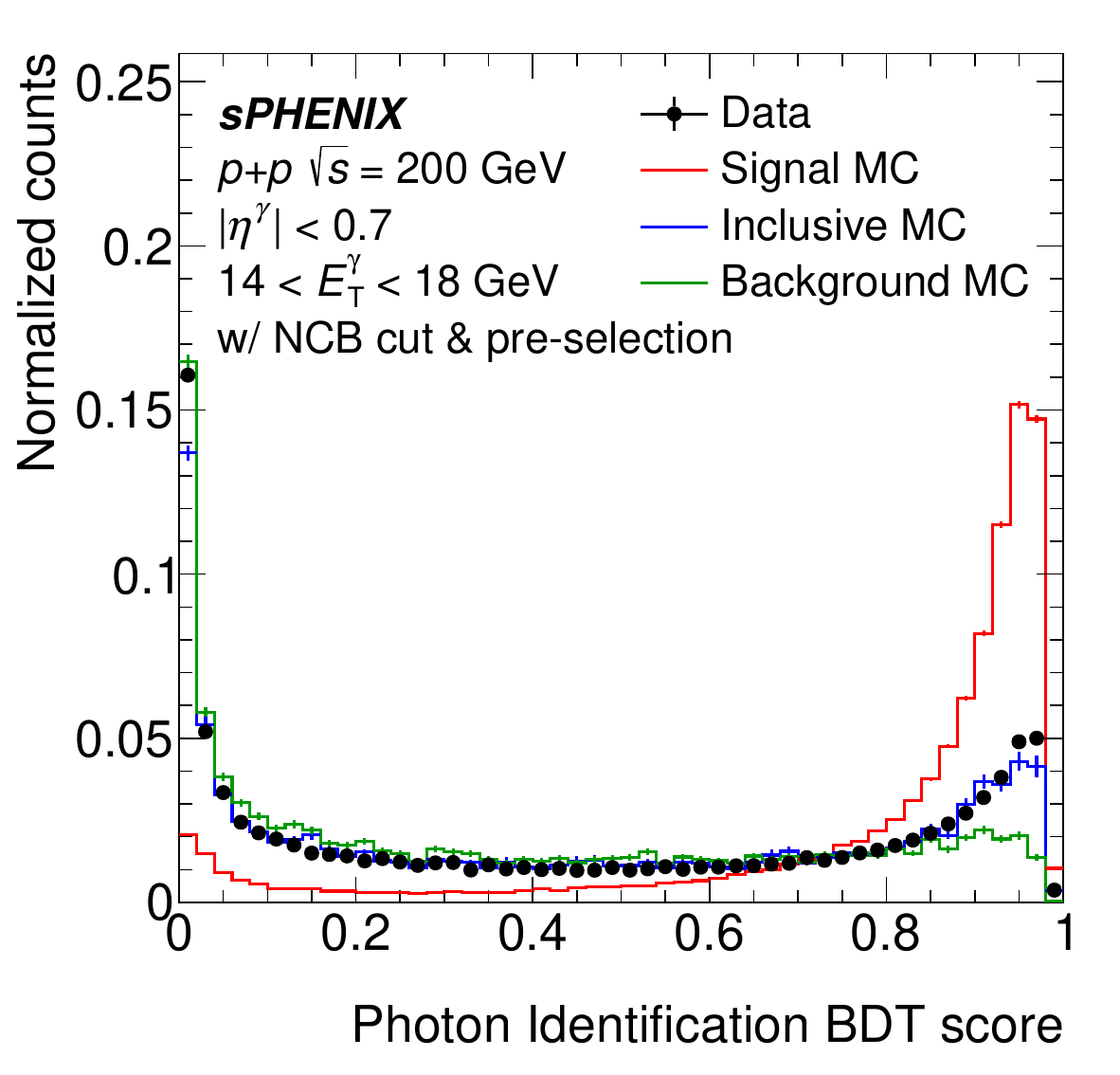}
    \caption{Distribution of the photon identification BDT score for data (black points), prompt photon signal MC (red), inclusive MC (blue), and background MC (green), after the NCB and pre-selection requirements, for $14<\etg<18~\GeV$ and $|\etag|<0.7$. The data and MC distributions are normalized such that the integral is unity.}
    \label{fig:bdt_dis}
\end{figure}

The distribution of the photon identification BDT score for $14<\etg<18~\GeV$ is shown in Figure~\ref{fig:bdt_dis}. The score is constructed to approach unity for prompt photon-like clusters and zero for background-like clusters: prompt photon signal MC accordingly populates the high-score region, while the inclusive MC peaks at low score. The small low-score excess in the signal MC arises from pile-up events whose shifted vertex distorts the reconstructed cluster kinematics and shower shape variable distributions. The data distribution is closer to the background MC except at high BDT score, reflecting the dominance of background photons in this kinematic region.

The ``tight'' photon identification (\gid) selection is defined by an \etg{}-dependent BDT-score threshold. The tight selection requires 
$\mathrm{BDT}_{\mathrm{ID}} > 0.8 - 0.00156\,(\etg - 10 \GeV)$, with $\etg$ in GeV. The ``non-tight'' \gid{} selection is defined by $0.6 - 0.01333\,(\etg - 10 \GeV) < \mathrm{BDT}_{\mathrm{ID}} < 0.7 + 0.00156\,(\etg - 10 \GeV)$, 
which enriches the background photon contribution and is used as the photon identification sideband in the data-driven purity calculation described in Sec.~\ref{ssec:abcd}. The $\mathrm{BDT}_{\mathrm{ID}}$ thresholds for both tight and non-tight regions used in this analysis are optimized to balance the statistics and purity that depend on \etg{}.

\subsection{Photon Isolation}
\label{ssec:iso}

The photon isolation requirement suppresses not only fragmentation photons, but also background photons from high-\pt neutral mesons decaying into two photons. Such neutral mesons are commonly produced inside jets and are therefore accompanied by additional particles near the photon candidate. In combination with the photon identification selection, the isolation requirement further reduces this background.

The reconstructed isolation transverse energy (\isoETreco) is computed from topologically reconstructed clusters (``topoclusters''), which provide resilience against calorimeter noise. The topoclusters are built across the EMCal, IHCal, and OHCal towers using the algorithm originally developed by ATLAS~\cite{ATLAS:2017topoclusters} and adapted to the sPHENIX three-layer calorimeter system.
The algorithm uses a $4$-$2$-$1$ significance configuration: cells with energy above $4\sigma$ of the per-cell noise level seed the clusters, neighboring cells above $2\sigma$ are added iteratively, and a final perimeter band above $1\sigma$ closes the clusters. The per-cell noise level is set to approximately three times the tower pedestal RMS. Local maxima above a layer-dependent threshold within a single topocluster are identified and used to split clusters containing more than one shower core, mitigating the merging of nearby showers. The \isoETreco is defined as the scalar \et sum of all topoclusters within $\DR=0.4$ around the photon direction, after subtracting the photon candidate's own cluster energy. The isolation requirement is chosen as $\isoETreco < 0.49 + 0.037\,\etg$ [GeV] and is tuned to maintain an approximately $80\%$ isolation efficiency. Non-isolated candidates used in the purity calculation are required to have \isoETreco{} at least $0.8$ GeV above the isolation threshold.

\begin{figure}[htbp!]
    \centering
    \includegraphics[width=0.45\linewidth]{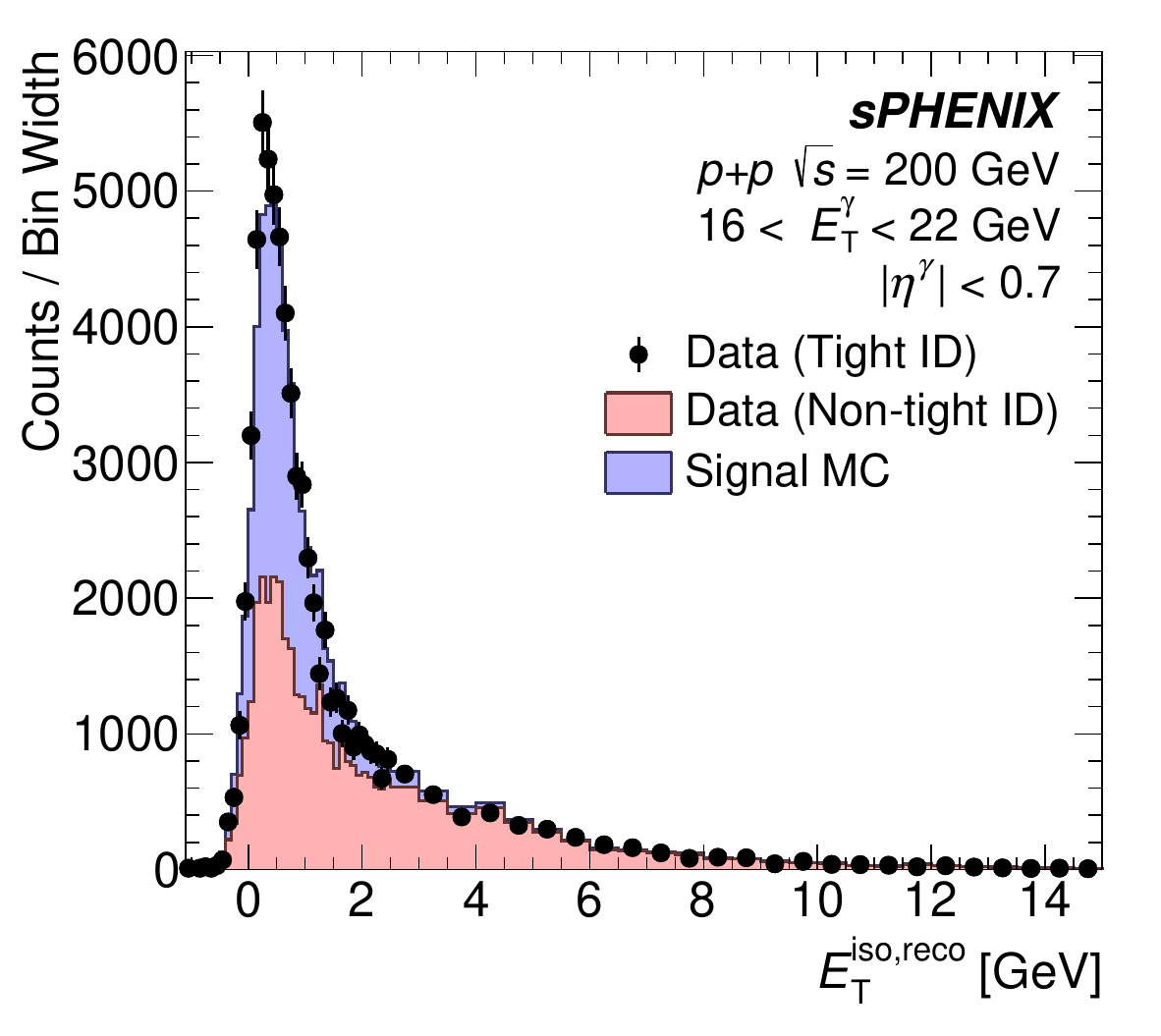}
    \caption{\isoETreco{} distributions of signal-enriched data with tight \gid, background-enriched data with non-tight \gid{} and signal MC with tight \gid. The background-enriched data and signal MC histograms are stacked.}
    \label{fig:isoDist}
\end{figure}

Figure~\ref{fig:isoDist} shows the \isoETreco{} distribution for tight \gid{} data, non-tight \gid{} data, and tight \gid{} signal MC. A small empirical correction is applied to the simulated \isoETreco{} to account for residual data-MC differences. The non-tight \gid{} data distribution is normalized to the tight \gid{} data in the background-dominated region, $\isoETreco>4~\GeV$, and is then combined with the tight \gid{} signal MC. The resulting sum reproduces the tight \gid{} data distribution at low \isoETreco{}, supporting the signal and background decomposition and demonstrating the difference in \isoETreco{} shape between the tight and non-tight selections.

\subsection{Signal Extraction}
\label{ssec:abcd}
\begin{figure}[htbp!]
    \centering
    \includegraphics[width=0.45\linewidth]{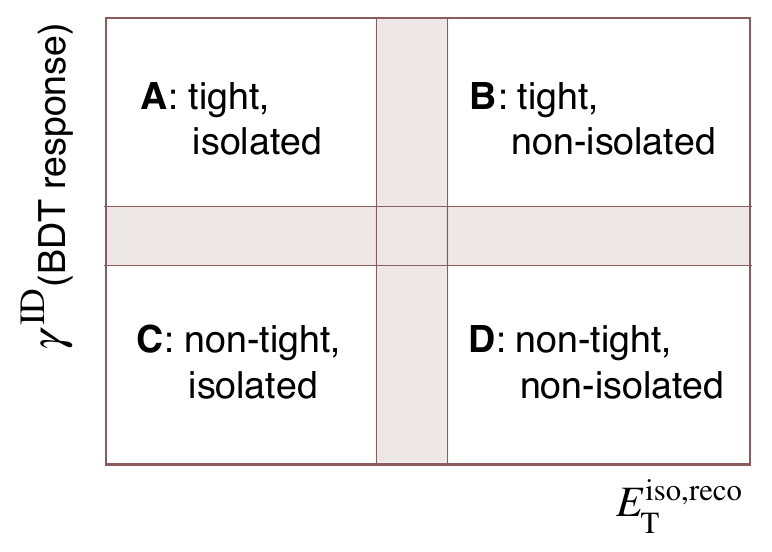}
    \caption{Diagram of the signal region (A) and sideband regions (B, C, D) used for purity estimation. The shaded bands indicate gaps between the signal and sideband regions.}
    \label{fig:sideband_diagram}
\end{figure}

The residual background remaining after the tight \gid{} and isolation requirements is estimated and statistically subtracted using a data-driven double-sideband technique~\cite{ATLAS:2016fta,ATLAS:2017nah}. The method uses four regions defined by the photon identification and isolation selections, as illustrated in Figure~\ref{fig:sideband_diagram}. Region A is the signal region, corresponding to tight \gid{} and isolated candidates. The three sideband regions are B (tight \gid{}, non-isolated), C (non-tight \gid{}, isolated) and D (non-tight \gid{}, non-isolated). These regions are defined using the tight and non-tight \gid{} selections described in Sec.~\ref{ssec:reco} and the isolated and non-isolated selections described in Sec.~\ref{ssec:iso}. Gaps are imposed between the signal and sideband regions to reduce the leakage of signal photons into the sidebands. 

The ratio of background yields in region C to region D is assumed to be the same as that in region A to region B. This assumption requires the photon identification and isolation selections to be independent for background candidates. For the dominant background from $\pi^0$ decays, the properties that determine the photon identification selection, such as the shower shape, energy asymmetry, and opening angle, are expected to be independent of the surrounding particle production, at fixed \et.  
To ensure the validity of this assumption, the \gid{} and pre-selection requirements are chosen to minimize their correlation with \isoET{} for background photons. Assuming that there is no leakage of the signal to the side-band regions (B, C, and D), the number of signal photons in region A is then calculated using:
\begin{equation}
N^{A}_{\text{signal}}
= N^{A}_{\text{raw}}
- N^{B}_{\text{raw}}
\left( \frac{N^{C}_{\text{raw}}}{N^{D}_{\text{raw}}} \right),
\label{eq:purity_noleakage}
\end{equation}
where $N^{A}_{\text{signal}}$ is the number of signal photons in region A, and $N^{X}_{\text{raw}}$ is the total photon candidate yield in region $X$, with $X = A, B, C,$ or $D$.

Equation~\ref{eq:purity_noleakage} is then modified to account for signal photons leaking into the sideband regions, using the truth-matched signal-photon ratios $f^{X,\text{MC}}$ between region X and region A determined in simulation:
\begin{equation}
N^{A}_{\text{signal}}
= N^{A}_{\text{raw}}
- \biggl[
  \bigl(
    N^{B}_{\text{raw}}
    - f^{B,\text{MC}}\,N^{A}_{\text{signal}}
  \bigr)
  \frac{
    \bigl(
      N^{C}_{\text{raw}}
      - f^{C,\text{MC}}\,N^{A}_{\text{signal}}
    \bigr)
  }{
    \bigl(
      N^{D}_{\text{raw}}
      - f^{D,\text{MC}}\,N^{A}_{\text{signal}}
    \bigr)
  }
\biggr],
\label{eq:purity_leak}
\end{equation}
The purity $\mathcal{P}$ is defined as the fraction of signal photons relative to the total photon candidates in the signal region A, $\mathcal{P} = N^{A}_{\text{signal}} / N^{A}_{\text{raw}}$.

The purity is calculated independently in each \etg{} bin. Its statistical uncertainty is estimated using a bootstrap procedure applied to $N^{A}_{\mathrm{signal}}$. The yields in regions A, B, C, and D are resampled $10\,000$ times assuming Poisson statistics, and the purity is recomputed for each pseudo-experiment. The statistical uncertainty in each \etg{} bin is taken as the standard deviation of the resulting purity distribution.

\begin{figure}[tbp!]
    \centering
    \includegraphics[width=0.45\linewidth]{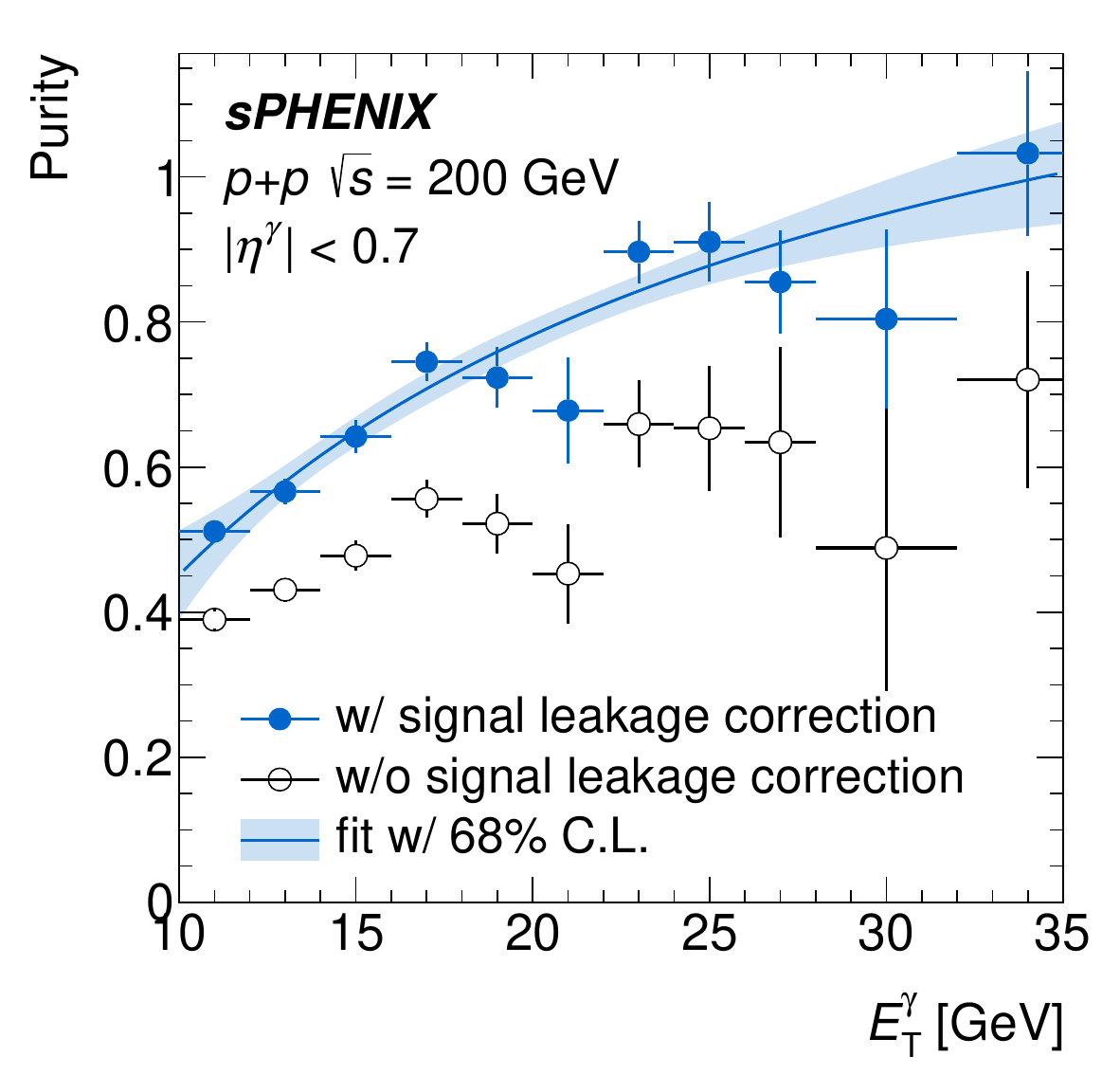}
    \caption{Purity as a function of \etg{} with and without signal leakage correction. The purity with leakage correction is fitted with a Pad\'e function, and the shaded area shows the $68$\% confidence interval of the fit.}    \label{fig:purity}
\end{figure}

Figure~\ref{fig:purity} shows the purity as a function of \etg{} with and without the signal leakage correction. The signal leakage correction increases the extracted purity, with the size of the correction ranging from approximately $0.1$ to $0.2$ depending on \etg{}. The truth-matched leakage fractions $f^{B,\text{MC}}$, $f^{C,\text{MC}}$, and $f^{D,\text{MC}}$ rise with \etg{} across the reported range, with $f^{B,\text{MC}}$ varying from approximately $8$\% to $13$\%, $f^{D,\text{MC}}$ from below $1$\% to $4$\%, and $f^{C,\text{MC}}$ rising from approximately $6$\% at low \etg{} to nearly $40$\% at the high end, where the non-tight isolated region (C) is the dominant source of signal leakage into the sidebands. The purity as a function of \etg{} is fitted with a smooth two-parameter function (a $[1/1]$ Pad\'e approximant), $P_{[1/1]}(x) = (a_0 + a_1 x)/(1 + b_1 x)$, to smooth statistical fluctuations in the bin-by-bin purity estimates. The fitted purity value in each \etg{} bin is then used to correct the photon yield passing the tight \gid{} and isolation requirements in data.


\subsection{Unfolding}
\label{ssec:unfolding}
The purity-corrected photon \etg{} yield distribution is unfolded using the D'Agostini Bayesian iterative method~\cite{DAgostini:2010hil} with the RooUnfold software package version 3.0.5~\cite{Adye:2011gm} to correct for detector effects on the energy response. The response matrix is constructed with reconstructed-\etg{} bins spanning $10$--$36\GeV{}$, extending beyond the fiducial cross section range of $12$--$32\GeV{}$ to account for bin migration. The truth-\etg{} axis is extended further, spanning $8$--$45\GeV{}$ to include overflow buffer bins.

To reduce sensitivity to the MC prior, the response matrix is reweighted using the ratio of the purity-corrected data yield to the MC signal photon yield. The unfolding is performed with two iterations, chosen to minimize the combined effect of iteration-dependent changes and statistical uncertainties. The unfolding correction changes the yield by approximately $5$--$10\%$, depending on \etg{}.

\subsection{Efficiency Correction}
\label{ssec:efficiency}
\begin{figure}[hbtp!]
    \centering
    \includegraphics[width=0.45\linewidth]{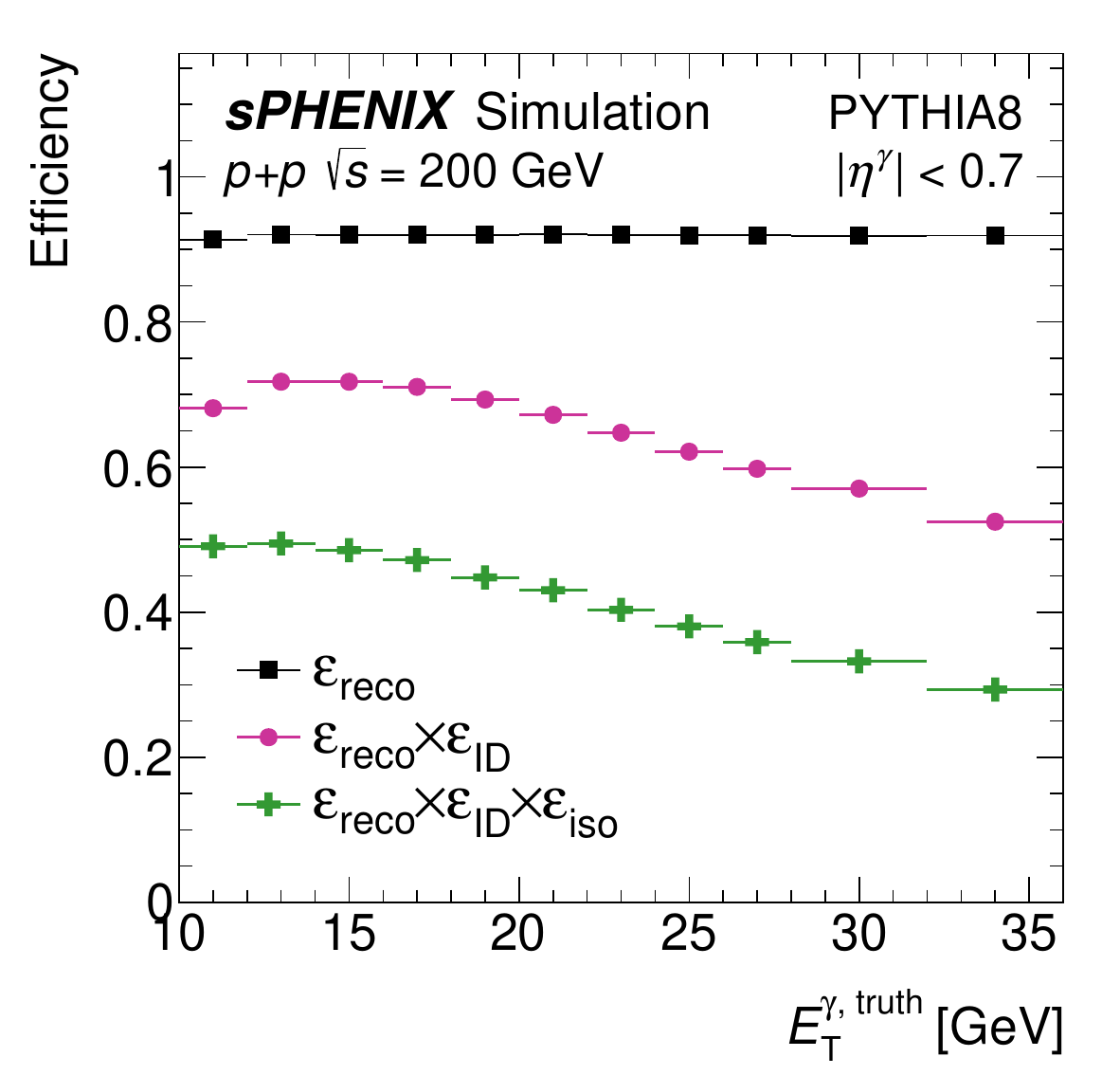}
    \caption{Reconstruction (\effreco), identification (\effid), isolation (\effiso), and combined (\efftot) efficiencies as a function of truth photon \etgtruth. The statistical uncertainties on the efficiencies are smaller than the marker size.}
    \label{fig:eff_photon}
\end{figure}

To account for losses of signal photons due to reconstruction and selection requirements, efficiencies estimated from MC simulation are applied as bin-by-bin corrections to the unfolded \etg{} spectrum, with the exception of the photon trigger efficiency.
The reconstruction efficiency, $\effreco$, is determined in simulation as the fraction of truth-level signal photons that are matched to a reconstructed EMCal cluster. It is approximately $92\%$ across the measured \etg{} range. The residual inefficiency comprises approximately $3.5\%$ from vertex-resolution-induced migration of the reconstructed cluster $\eta$ across the fiducial $|\etag|<0.7$ boundary, approximately $1.8\%$ from the tower acceptance (inactive or masked towers in the EMCal), and the remainder from photon conversions and other effects. The identification efficiency, $\effid$, and isolation efficiency, $\effiso$, are defined as conditional efficiencies for truth-matched reconstructed photons to satisfy the tight \gid{} and reconstruction-level isolation requirements, respectively, where the isolation efficiency corrects the reconstruction-level isolation yield back to the truth-isolated fiducial, absorbing both the cone-size and threshold differences between the two definitions. The reconstruction-level cone size $\Delta R = 0.4$ is chosen to be larger than the truth-level $\Delta R = 0.3$ to improve experimental performance. Figure~\ref{fig:eff_photon} shows the efficiencies as a function of \etgtruth for the individual contributions from $\effreco$, $\effid$, and $\effiso$, together with their combined efficiency.

The MBD-vertex efficiency, $\varepsilon_{\mathrm{vtx}}$, is defined as the fraction of truth-level signal events in which the MBD records at least one hit on each side and yields a reconstructed vertex within $|z_\mathrm{reco}|<60~\mathrm{cm}$. This efficiency decreases from approximately $59\%$ at $\etg\!\simeq\!12$~\GeV{} to approximately $52\%$ at $\etg\!\simeq\!30$~\GeV. The $z_\mathrm{reco}$ requirement produces $62-63$\% of this inefficiency and the rest is a result of the hit requirement. The photon trigger efficiency is at least $99.5\%$ for $\etg>8.5\GeV{}$ and is applied per cluster before the remaining efficiency corrections, as described in Sec.~\ref{ssec:eventsel}. 

\subsection{Cross Section Determination}
\label{ssec:xsec}

The \etg{}-differential cross section of isolated prompt photons is calculated as
\[
\frac{d^{2}\sigma}{dE_\text{T}^\gamma d\eta}
=
\frac{1}{\mathscr{L}}
\frac{Y^\mathrm{rec}}{\mathcal{E}\Delta\etg{}\Delta\etag{}} ,
\]
where $\mathcal{E}$ is the combined efficiency for photon reconstruction, identification, isolation, trigger, and MBD-vertex, $\Delta\etg$ and $\Delta\etag$ are the bin widths.
$Y^\mathrm{rec}$ is the purity-corrected and unfolded photon yield. The integrated luminosity, $\mathscr{L}=64.4^{+5.9}_{-4.3}~\pb$, is determined from the minimum-bias trigger cross section measured in a Vernier scan and the number of minimum-bias-triggered events in the analyzed data sample, with a per-run pile-up correction derived from the measured $\mu$. 

\section{Systematic Uncertainties}
\label{sec:systematics}
The systematic uncertainties are divided into the following sources: photon energy scale and resolution, purity, efficiency, unfolding, pile-up, NCB rejection, and luminosity. Each source is evaluated by repeating the full analysis with the corresponding variation. For each \etg bin, relative deviations from the nominal cross section are added in quadrature to give the total systematic uncertainty. 

The reconstructed EMCal cluster transverse energy in MC is varied by a multiplicative factor of $\pm 1.5\%$ to account for the energy scale calibration uncertainty. An additional uncertainty is assigned for the residual energy scale non-linearity associated with the EMCal response effects described in Sec.~\ref{sec:data_sim}. The residual data-MC discrepancy in this non-linearity, constrained by test-beam measurements~\cite{Aidala:2020toz}, is propagated as a systematic uncertainty. For the EMCal energy resolution uncertainty, two variations that bracket the nominal extra smearing from below and above are taken: a no-extra-smearing variation in which $\sigma_{+MC}\!=\!0$, and a wider variation $\sigma_{+MC}\!\approx\!6\%$ approximately flat in \etgtruth.

The systematic uncertainty on the purity is evaluated from six contributions, which are added in quadrature. Three contributions arise from variations of the sideband definitions: the tight \gid{} threshold, the non-tight \gid{} threshold, and the non-isolation boundary. Two contributions are associated with the purity fit: the nominal $[1/1]$ Pad\'e approximant is replaced with an error-function alternative to probe the functional-form dependence, and the nominal fit parameters are shifted by their $\pm 1\sigma$ uncertainties to account for the statistical precision of the fit. In the highest-\etg{} bin, a small number of systematic variations result in fitted purity values slightly above unity. These values are capped at unity to maintain the physical constraint on the purity, with a negligible impact on the systematic uncertainty.
The sixth contribution is an MC-closure uncertainty. It is determined by comparing the truth-level purity in MC with the purity extracted from reconstructed level simulation by applying the same sideband method as in data. This uncertainty accounts for residual correlations between the photon identification and isolation requirements for background candidates. The MC-closure contribution dominates the purity uncertainty at low \etg{}.

The remaining systematic uncertainties are evaluated as follows. The isolation efficiency uncertainty is taken from variations of the empirical \isoET{} correction described in Sec.~\ref{ssec:reco} and from the comparison of the isolation efficiency between \pythia{} and \herwig{} (Sec.~\ref{ssec:mc}), together with a separate envelope assigned to the MBD-vertex efficiency based on data-driven studies.  These studies compared the fraction of dijet events with a reconstructed MBD vertex in data and simulation, and assigned the observed difference as a systematic uncertainty.
The unfolding uncertainty is evaluated by repeating the unfolding without the prior reweighting and by varying the number of Bayesian iterations. The pile-up mixture fraction in MC is varied according to alternative values obtained from fits to the shower shape comparisons between data and MC. The non-collision-background uncertainty is evaluated by varying the NCB BDT score requirement. Finally, the luminosity uncertainty of $\mathscr{L}=64.4^{+5.9}_{-4.3}~\mathrm{pb}^{-1}$ is propagated from the in-situ Vernier scan measurement of the MBD inelastic cross section and is treated as fully correlated across all \etg{} bins. These contributions are subdominant compared to the leading systematic uncertainties.

The systematic variations are treated according to whether they provide separate up and down deviations (two-sided), a single deviation (one-sided), or multiple alternative deviations from the nominal result. Two-sided sources (energy scale, energy resolution, non-isolation boundary, NCB cut, purity $68\%$ confidence interval, luminosity) yield asymmetric up and down deviations that are kept separate per \etg{} bin. One-sided sources (tight \gid{} threshold, non-tight \gid{} threshold, fit functional form, MC closure correction, isolation efficiency, unfolding reweighting, pile-up fraction) yield a single deviation that is symmetrized about the nominal. The unfolding iteration uncertainty is evaluated using two variations (three and four iterations, with two as nominal), and the maximum deviation in each bin is symmetrized about the nominal result.

\begin{figure}[tbp!]
    \centering
     \includegraphics[width=0.8\linewidth]{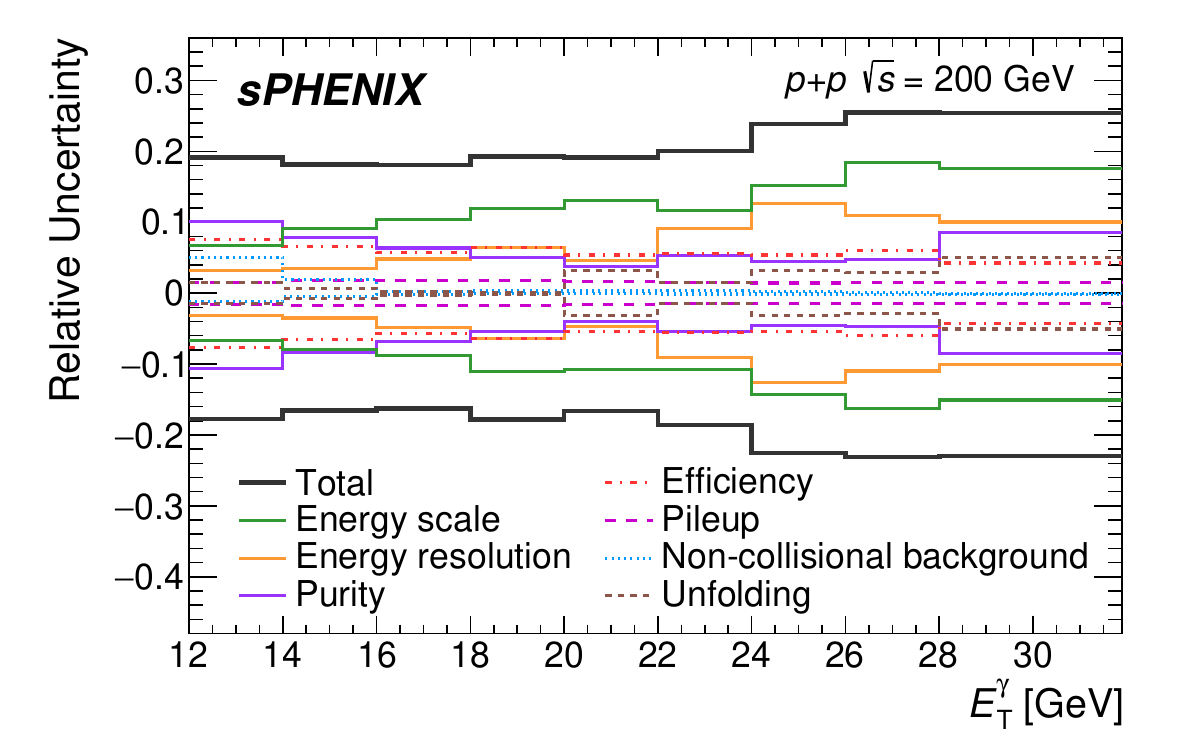}
    \caption{Breakdown of relative systematic uncertainties as a function of \etg. The total envelope includes the global luminosity uncertainty (added in quadrature, \etg-independent at $^{+9.1\%}_{-6.8\%}$) in addition to the seven components shown.}    \label{fig:syst_sum}
\end{figure}

The total systematic uncertainty and its individual contributions are shown in Figure~\ref{fig:syst_sum}. The energy-scale, energy-resolution, and purity-closure uncertainties, which are the three largest contributions, are each strongly correlated across \etg{} bins. The energy scale and energy resolution uncertainties dominate the total uncertainty at high \etg{} while the purity closure uncertainty dominates at low \etg{}. The total systematic uncertainty is obtained by adding all components in quadrature, and reaches approximately $-23\%/+26\%$ in the highest \etg{} region.

\section{Results}
\label{sec:results}

\begin{figure}
    \centering
    \includegraphics[width=0.75\linewidth]{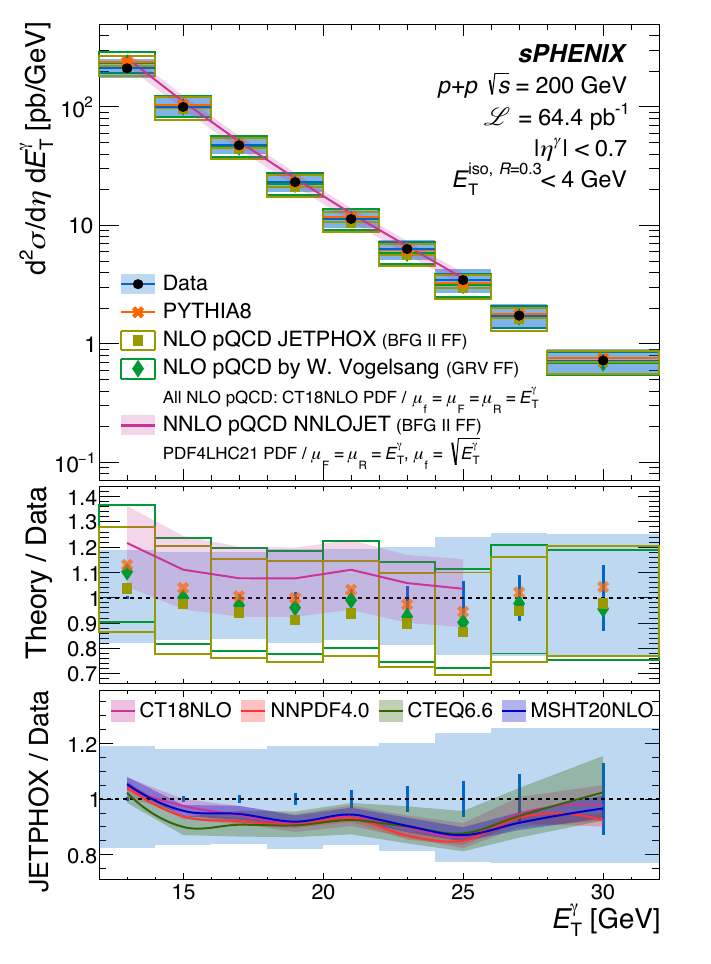}
    \caption{Differential cross section of isolated prompt photons as a function of \etg{} in \pp{} collisions at \comHEP{}, in the kinematic range $|\etag|<0.7$ and $12<\etg<32~\GeV$. For the data, statistical uncertainties are shown as vertical bars, and the total systematic uncertainty as shaded bands. The data are compared in the upper panel with predictions from \pythia{}~8.307 (Detroit tune), \jetphox{}, the NLO pQCD calculation by Vogelsang, and the NNLO pQCD calculation from \textsc{NNLOJET}, all evaluated with the same truth-level isolation requirement as the data. 
    The boxes around the NLO pQCD predictions and the shaded band around the NNLO pQCD prediction show the scale-variation uncertainties. The middle panel shows the theory-to-data ratio with the experimental statistical and systematic uncertainties drawn around unity as vertical bars and a shaded band, respectively. The lower panel represents the \jetphox{} predictions with four different proton PDF sets with the PDF uncertainties shown as shaded bands around the lines  (\texttt{CT18NLO}~\cite{Hou:2019efy}, \texttt{NNPDF4.0}~\cite{NNPDF:2021njg}, \texttt{CTEQ6.6}~\cite{Nadolsky:2008zw}, and \texttt{MSHT20NLO}~\cite{Bailey:2020ooq}) divided by the data.
    }   \label{fig:final}
\end{figure}

The differential cross section of isolated prompt photons as a function of \etg{} in $|\etag|<0.7$ is shown in Figure~\ref{fig:final}. The measured cross section spans approximately two orders of magnitude over the reported \etg{} range. The result is compared with predictions from the \pythia{}~$8.307$ MC generator with the Detroit tune~\cite{Aguilar:2021sfa}, the NLO pQCD calculation from \jetphox{}~v$1.3.1\_4$~\cite{Aurenche:2006vj}, the NLO pQCD calculation by Vogelsang~\cite{Gordon:1993qc} and the NNLO pQCD predictions from \nnlojet{}~\cite{NNLOJET:2025rno}.  All predictions are evaluated with the same particle-level isolation requirement (\isoET{} $< 4~\GeV{}$ within $\DR=0.3$) that the data are corrected to.

The \jetphox{} and Vogelsang NLO calculations use the \texttt{CT18NLO}~\cite{Hou:2019efy} as their nominal PDF set. The \jetphox{} prediction includes the BFG~set~II~\cite{Bourhis:1997yu} parton-to-photon fragmentation functions, while the Vogelsang calculation uses the GRV~\cite{Gluck:1992zx} fragmentation functions. For both NLO pQCD calculations, the renormalization, factorization, and fragmentation scales are set to $\mu_{R}=\mu_{F}=\mu_{f}=\etg$, and the scale uncertainty is obtained by simultaneously varying these scales by factors of $1/2$ and $2$. 

The \nnlojet{} predictions are evaluated using the \texttt{PDF4LHC21}~\cite{PDF4LHCWorkingGroup:2022cjn} PDF set and the BFG~set~II fragmentation functions. The renormalization and initial-state factorization scales are set to $\mu_R=\mu_F=\etg$, while the fragmentation scale is set to $\mu_f=\sqrt{\etg}$ following the prescription of Ref.~\cite{Chen:2022gpk}. The theory uncertainty is evaluated using a 15-point variation of $\mu_R$, $\mu_F$, and $\mu_f$. The integration statistical uncertainty of the theory calculation is negligible compared with the scale variation uncertainty.

As shown in the ratio in the middle panel of Fig.~\ref{fig:final}, the \pythia{} and NLO pQCD predictions are consistent with the measurement within the quoted uncertainties. The NNLO correction from \nnlojet{} increases the corresponding NLO prediction by an approximately \etg{}-independent factor of 1.2, and the NNLO prediction is also consistent with the measurement within the quoted uncertainties. The lower panel shows the \jetphox{} predictions obtained with several PDF sets, \texttt{CT18NLO}~\cite{Hou:2019efy}, \texttt{NNPDF4.0}~\cite{NNPDF:2021njg}, \texttt{CTEQ6.6}~\cite{Nadolsky:2008zw}, and \texttt{MSHT20NLO}~\cite{Bailey:2020ooq}, together with their PDF uncertainties, illustrating the sensitivity of the prediction to the proton PDF in the probed kinematic range.

\begin{figure}
    \centering
    \includegraphics[width=0.75\linewidth]{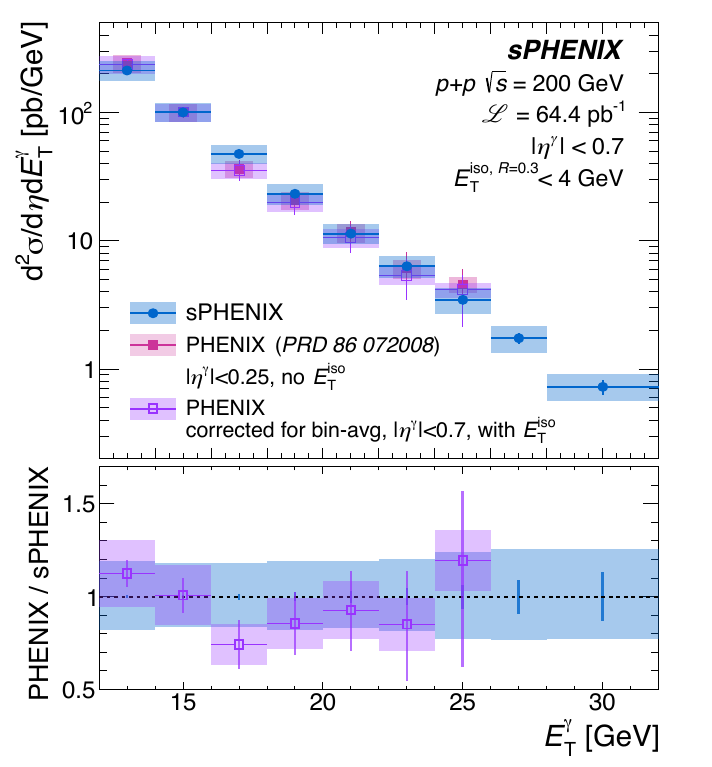}
    \caption{Comparison of the present isolated prompt photon cross section measurement (blue) with the PHENIX measurement~\cite{PHENIX:2012jgx} (pink). The PHENIX measurement is corrected to match to the sPHENIX bin-width, pseudorapidity acceptance $|\eta^\gamma|<0.7$, and isolated photon definition, and shown as purple open squares. Details of the correction procedure are described in the text. Statistical uncertainties are shown as vertical bars, and total systematic uncertainties are shown as shaded bands. The bottom panel shows the ratio of the corrected PHENIX data to the sPHENIX measurement, with the vertical bar at each ratio point representing the statistical uncertainty of the PHENIX measurement. The sPHENIX statistical (vertical bars) and total systematic (shaded band) uncertainties are drawn around unity across the full sPHENIX \etg range.
    }
    \label{fig:final_phenix}
\end{figure}

The cross section is compared with the measurement by the PHENIX Collaboration~\cite{PHENIX:2012jgx} in Fig.~\ref{fig:final_phenix}. The PHENIX measurement was performed without an isolation requirement, in a narrower acceptance of $|\eta^\gamma|<0.25$, and reports the cross section at the bin centers rather than as bin-integrated values. To enable a direct comparison, three corrections are applied to the published PHENIX points. First, a bin-width correction is applied to convert the bin-center cross section to the corresponding bin-integrated value, using a fit to the PHENIX spectrum. Second, a pseudorapidity-density rescaling is applied to account for the different rapidity acceptances. This correction divides each PHENIX point by the ratio $R_{\eta}(\etg)$ of the prompt photon truth density in $|\eta^\gamma|<0.25$ to that in $|\eta^\gamma|<0.7$, evaluated with \jetphox{}. The ratio $R_{\eta}(\etg)$ rises from approximately unity at $\etg=12~\GeV{}$ to approximately $1.15$ at $\etg=32~\GeV{}$, reflecting the narrowing of the rapidity distribution for $2\to2$ prompt photon production at higher Bjorken-$x$. Third, an isolation correction is applied to bring the inclusive PHENIX measurement to the isolated definition used in this analysis, computed as the ratio of the \jetphox{} NLO pQCD predictions with and without the isolation requirement. The sPHENIX measurement and the corrected PHENIX measurement agree within the quoted experimental uncertainties.
The sPHENIX measurement provides coverage over a larger \etg{} range and rapidity acceptance.

\begin{figure}
    \centering
    \includegraphics[width=0.65\linewidth]{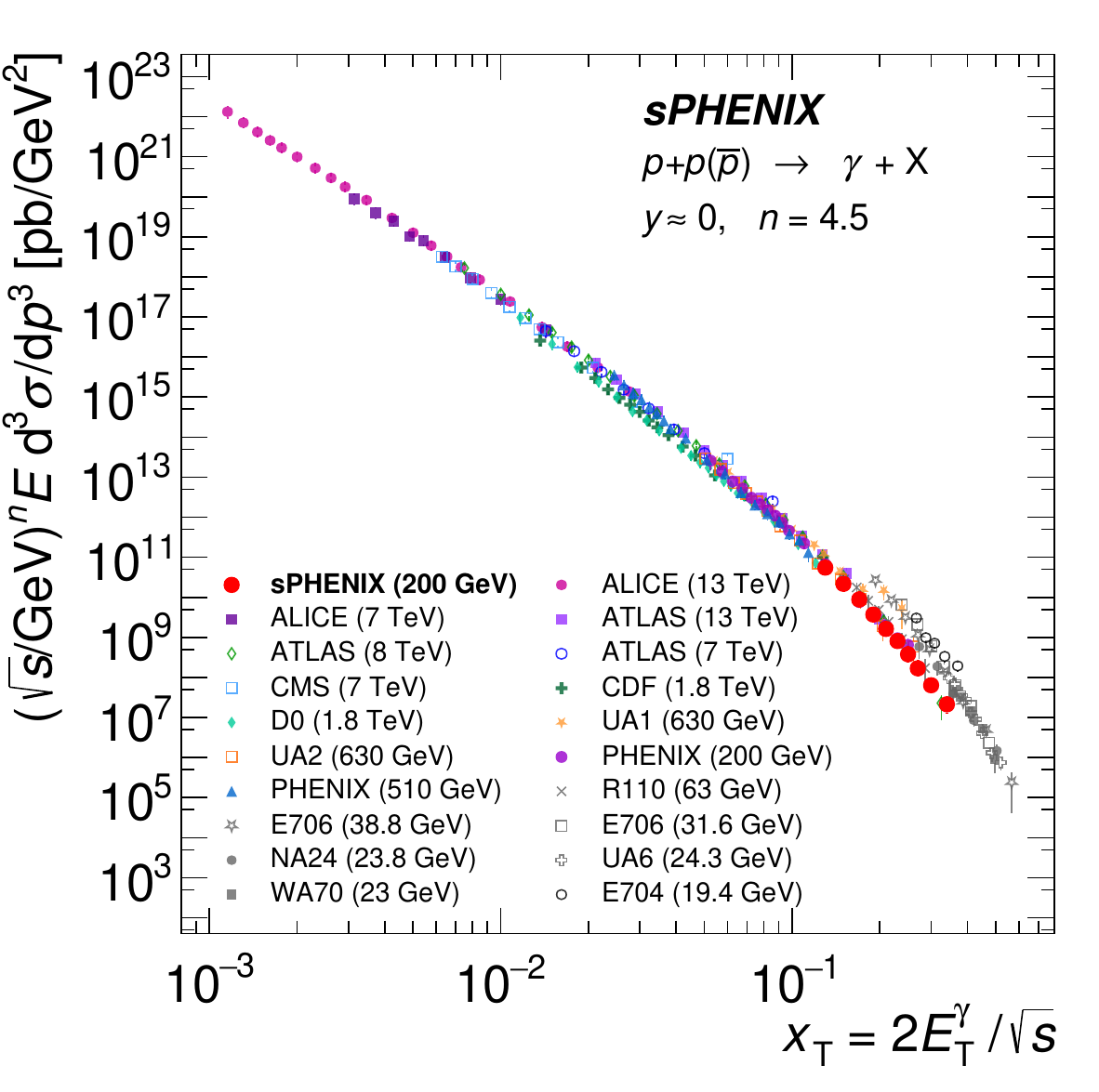}
    \caption{\xt scaling of the invariant cross section for prompt photon production in \pp{} and $p+\bar{p}$ collisions from various measurements at different collision energies. The vertical axis is $(\sqrt{s}/\GeV)^{n}\,E\,d^3\sigma/dp^3$ with $n=4.5$; the horizontal axis is $\xt = 2\etg/\sqrt{s}$. The sPHENIX measurement of this paper (red full circle) is overlaid on prompt photon measurements at the LHC, Tevatron, SppS, RHIC, ISR, and fixed-target experiments. 
    }
    \label{fig:xtscaling}
\end{figure}

The present measurement is further compared with the world compilation of prompt photon cross section measurements through \xt scaling~\cite{Bock:2019rqx}. In this representation, the invariant cross section $E\,d^3\sigma/dp^3$ is multiplied by $(\sqrt{s})^{n}$ with $n = 4.5$ and plotted versus $\xt = 2\etg/\sqrt{s}$, which at leading order in pQCD collapses measurements at different $\sqrt{s}$ onto a single universal curve. Figure~\ref{fig:xtscaling} shows the present result overlaid on prompt photon measurements at the LHC~\cite{ATLAS:2011ezy,ATLAS:2016fta,ATLAS:2017nah,CMS:2010svd}, Tevatron~\cite{CDF:1994wme,D0:1996wgi}, SppS~\cite{UA1:1988zam,UA2:1992yma}, RHIC~\cite{PHENIX:2012jgx,PHENIX:2022lgn}, ISR~\cite{CMOR:1989qzc}, and fixed-target~\cite{NA24:1986cpj,WA70:1987vvj,UA6:1998ozg,E704:1995her,FermilabE706:2004emk} experiments spanning $\sqrt{s} = 19.4~\GeV$ to $13$ TeV. The sPHENIX measurement lies on the universal curve and is consistent with the PHENIX $200$~\GeV{} measurement within the combined uncertainties, providing an external cross-experiment check on the absolute normalization of this measurement. Some fixed-target measurements at the lowest $\sqrt{s}$ deviate above the universal curve at high \xt, consistent with the exception at low $\sqrt{s}$ noted previously~\cite{PHENIX:2012jgx}.

\clearpage
\pagebreak

\section{Summary}
\label{sec:summary}

The differential cross section of isolated prompt photon production as a function of transverse energy, \etg{}, has been measured in proton-proton collisions at \comHEP{}, corresponding to an integrated luminosity of
$\mathscr{L}=64.4~\pb$. The measurement is reported in the kinematic range $|\etag|<0.7$ and $12<\etg<32\GeV{}$, with an isolation requirement of $\isoET<4~\GeV{}$ within $\DR=0.3$.

Photon candidates are reconstructed in the electromagnetic calorimeter and identified using a boosted decision tree based on electromagnetic shower shape observables. Isolated photons are selected using an isolation transverse energy computed with both the electromagnetic and hadronic calorimeters. Residual background photons are statistically subtracted using a data-driven double-sideband technique. The photon yield is corrected for purity and efficiency and unfolded for detector response using the iterative Bayesian method.

The next-to-leading-order pQCD predictions from \jetphox{} and Vogelsang, the next-to-next-to-leading-order pQCD prediction from \nnlojet{}, and the \pythia{} MC generator using the Detroit tune, are consistent with the measured cross section. Comparisons with different parton distribution function sets are also presented. The data are further compared with the PHENIX measurement at the same collision energy~\cite{PHENIX:2012jgx}. This measurement provides coverage over a higher \etg{} range and a larger rapidity acceptance than the previous PHENIX result, with substantially improved statistical precision. The result provides a test of pQCD and a \pp{} baseline for measurements of photon and photon+jet observables in heavy-ion collisions.



\section*{Acknowledgements}

We thank the Collider-Accelerator Division, Scientific Computing and Data Facilities (SCDF) and Physics Departments at Brookhaven National Laboratory and the staff of the other sPHENIX participating institutions for their vital contributions. We acknowledge support from the Office of Nuclear Physics and Graduate Student Research (SCGSR) program in the Office of Science of the U.S. Department  of Energy, the U.S. National Science Foundation, the National Science and Technology Council, the Ministry of Education of Taiwan, and the Ministry of Economic Affairs (Taiwan), the Ministry of Education, Culture, Sports, Science, and Technology and the  Japan Society for the Promotion of Science (Japan), Basic Science Research Programs through NRF funded by the Ministry of Education and the Ministry of Science and ICT (Korea) and the Swedish Research Council, VR (Sweden).

\bibliographystyle{unsrturl}
\bibliography{references}

\newpage

\appendix

\section*{The sPHENIX Collaboration}

\begin{flushleft}
\small

M.~I.~Abdulhamid$^{12}$,
U.~Acharya\,\orcidlink{0000-0001-8560-963X}\,$^{12}$,
G.~Adawi$^{12}$,
I.~Ahmed\,\orcidlink{0000-0003-4642-5023}\,$^{43}$,
C.~A.~Aidala\,\orcidlink{0000-0001-9540-4988}\,$^{24}$,
Y.~Akiba$^{37}$,
M.~Alfred$^{13}$,
A.~Alsayegh$^{9}$,
D.~M.~Anderson\,\orcidlink{0000-0003-3845-2304}\,$^{16}$,
V.~V.~Andrieux\,\orcidlink{0000-0001-9957-9910}\,$^{14}$
% ...

\section*{Collaboration Institutes}

$^{1}$ Augustana University, Sioux Falls, South Dakota\\
$^{2}$ Baruch College, City University of New York, New York, New York\\
$^{3}$ Brookhaven National Laboratory, Upton, New York\\
$^{4}$ University of California, Los Angeles, California\\
$^{5}$ Universit\'{e} Paris-Saclay --- CEA --- IRFU, Gif-sur-Yvette, France\\
$^{6}$ University of Colorado, Boulder, Colorado\\
$^{7}$ Columbia University, New York, New York\\
$^{8}$ Debrecen University, Debrecen, Hungary\\
$^{9}$ Florida Agricultural and Mechanical University, Tallahassee, Florida\\
$^{10}$ Florida State University, Tallahassee, Florida\\
$^{11}$ Fudan University, Shanghai\\
$^{12}$ Georgia State University, Atlanta, Georgia\\
$^{13}$ Howard University, Washington, District of Columbia\\
$^{14}$ University of Illinois at Urbana-Champaign, Urbana, Illinois\\
$^{15}$ Institute for Information Industry, Taipei\\
$^{16}$ Iowa State University, Ames, Iowa\\
$^{17}$ Japan Atomic Energy Agency, Naka, Ibaraki, Japan\\
$^{18}$ Korea University, Seoul, Korea\\
$^{19}$ Lawrence Berkeley National Laboratory, Berkeley, California\\
$^{20}$ Lawrence Livermore National Laboratory, Livermore, California\\
$^{21}$ Lehigh University, Bethlehem, Pennsylvania\\
$^{22}$ Los Alamos National Laboratory, Los Alamos, New Mexico\\
$^{23}$ Massachusetts Institute of Technology, Cambridge, Massachusetts\\
$^{24}$ University of Michigan, Ann Arbor, Michigan\\
$^{25}$ Morgan State University, Baltimore, Maryland\\
$^{26}$ Muhlenberg College, Allentown, Pennsylvania\\
$^{27}$ Nara Women's University, Nara, Nara, Japan\\
$^{28}$ National Central University, Taoyuan City\\
$^{29}$ National Research Nuclear University, MEPhI, Moscow Engineering Physics Institute, Moscow, Russia\\
$^{30}$ National Taiwan University, Taipei\\
$^{31}$ University of North Carolina, Greensboro, North Carolina\\
$^{32}$ Oak Ridge National Laboratory, Oak Ridge, Tennessee\\
$^{33}$ Ohio University, Athens, Ohio\\
$^{34}$ Peking University, Beijing\\
$^{35}$ Purdue University, West Lafayette, Indiana\\
$^{36}$ Pusan National University, Pusan, Korea\\
$^{37}$ RIKEN Nishina Center for Accelerator-Based Science, Wako, Saitama, Japan\\
$^{38}$ Rikkyo University, Toshima, Tokyo, Japan\\
$^{39}$ Rutgers University, Piscataway, New Jersey\\
$^{40}$ State University of New York, Stony Brook, New York\\
$^{41}$ Sun Yat-sen University, Guangzhou, Guangdong\\
$^{42}$ Instituto de Alta Investigaci\'{o}n, Universidad de Tarapac\'{a}, Arica, Chile\\
$^{43}$ University of Tennessee, Knoxville, Tennessee\\
$^{44}$ United States Merchant Marine Academy, Kings Point, New York\\
$^{45}$ Vanderbilt University, Nashville, Tennessee\\
$^{46}$ Wayne State University, Detroit, Michigan\\
$^{47}$ Weizmann Institute of Science, Rehovot, Israel\\

\end{flushleft}

\end{document}